\documentclass[conference,compsoc]{IEEEtran}
\IEEEoverridecommandlockouts
\usepackage{cite}
\usepackage{amsmath,amssymb,amsfonts}
\usepackage{algorithmic}
\usepackage{graphicx}
\usepackage[table,xcdraw]{xcolor}
\usepackage[font=small]{caption,subcaption}
\usepackage{wrapfig}
\usepackage{textcomp}
\usepackage{xcolor}
\def\BibTeX{{\rm B\kern-.05em{\sc i\kern-.025em b}\kern-.08em
    T\kern-.1667em\lower.7ex\hbox{E}\kern-.125emX}}
\usepackage{url}
\usepackage{booktabs} 
\usepackage{multirow}
\usepackage[inline]{enumitem}

\usepackage{color,soul}
\usepackage{xcolor}
\usepackage{titlesec}
\usepackage{cleveref}

\titlespacing\section{0pt}{12pt plus 4pt minus 2pt}{0pt plus 2pt minus 2pt}
\titlespacing\subsection{0pt}{12pt plus 4pt minus 2pt}{0pt plus 2pt minus 2pt}
\titlespacing\subsubsection{0pt}{12pt plus 4pt minus 2pt}{0pt plus 2pt minus 2pt}

\begin{document}

\title{Blockchain Interoperability Landscape
}

\author{\IEEEauthorblockN{Inwon Kang}
\IEEEauthorblockA{\textit{Department of Computer Science} \\
\textit{Rensselaer Polytechnic Institute}\\
Troy, NY, USA \\
kangi@rpi.edu}
\and
\IEEEauthorblockN{Aparna Gupta}
\IEEEauthorblockA{\textit{Lally School of Management} \\
\textit{Rensselaer Polytechnic Institute}\\
Troy, NY, USA \\
guptaa@rpi.edu}
\and
\IEEEauthorblockN{Oshani Seneviratne}
\IEEEauthorblockA{\textit{Department of Computer Science} \\
\textit{Rensselaer Polytechnic Institute}\\
Troy, NY, USA \\
senevo@rpi.edu}
}

\IEEEoverridecommandlockouts
\IEEEpubid{\makebox[\columnwidth]{978-1-6654-8045-1/22/\$31.00~\copyright2022 IEEE \hfill} \hspace{\columnsep}\makebox[\columnwidth]{ }}
\maketitle
\IEEEpubidadjcol


\begin{abstract}
Blockchain has become a popular emergent technology in many industries. It is suitable for a broad range of applications, from its base role as an immutable distributed ledger to the deployment of distributed applications. Many organizations are adopting the technology, but choosing a specific blockchain implementation in an emerging field exposes them to significant technology risk. Selecting the wrong implementation could expose an organization to security vulnerabilities, reduce access to its target audience, or cause issues in the future when switching to a more mature protocol. Blockchain interoperability aims to solve this adaptability problem by increasing the extensibility of blockchain, enabling the addition of new use cases and features without sacrificing the performance of the original blockchain. 
However, most existing blockchain platforms need to be designed for interoperability, and simple operations like sending assets across platforms create problems.  
Cryptographic protocols that are secure in isolation may become insecure when several different (individually secure) protocols are composed. 
Similarly, utilizing trusted custodians may undercut most of the benefits of decentralization offered by blockchain-based systems.
Even though there is some research and development in the field of blockchain interoperability, a characterization of the interoperability solutions for various infrastructure options is lacking. This paper presents a methodology for characterizing blockchain interoperability solutions that will help focus on new developments and evaluate existing and future solutions in this space.
\end{abstract}

\begin{IEEEkeywords}
blockchain interoperability, survey, solution characterization
\end{IEEEkeywords}

\section{Introduction}

Since the introduction of Bitcoin~\cite{nakamoto2008bitcoin}, blockchain technology has continued to evolve and be applied to fields beyond digital asset management. The introduction of Ethereum and its programmability~\cite{wood2014ethereum} has supported impressive research and development in the blockchain space toward new mechanisms and projects to extend its use cases. However, many blockchains to date suffer from a lack of interoperability, which limits their scope of use cases. As projects mature, their blockchain interoperability emerges as an increasingly important feature. Without standardized protocols, projects seeking blockchain interoperability resort to their unique architecture and mechanisms for a specific purpose, and designing a universally interoperable framework requires understanding the goals of past projects and how they sought to achieve interoperability. This paper characterizes the blockchain interoperability landscape by examining several existing projects and comparing their similarities and differences. 

\subsection{What is Blockchain Interoperability?}
We define blockchain interoperability as the ability of blockchains to work together rather than being limited to their isolated environments. In this regard, interoperability for blockchain-based systems can mean different things, such as supporting asset exchange, asset transfer, or data exchange. Bitcoin, the first cryptocurrency to gain widespread attention, was not designed for interoperability or programmability. However, the emergence of Ethereum, followed by many other blockchains, has introduced many new constructs to the blockchain world. For instance, Ethereum introduced programmable smart contracts and the ability to send arbitrary data.

The ability of blockchain to handle more than just transactions has extended the possibilities of this technology. Having different blockchains, each with its niche, bring several benefits. However, it also raises questions about their ability to work together. Therefore, blockchain interoperability is a challenging goal to achieve in the current landscape. Blockchain technologies are inherently ``trustless,'' meaning parties without previously established relationships can interact reliably. However, no system works in isolation, and these blockchain primitives must be extended to the data coming from beyond a single blockchain ecosystem. This challenge is known as the ``oracle problem''~\cite{caldarelli2020understanding}. In other words, a blockchain must devise a verification approach for the data received from outside its ecosystem. 

Another question faced is how new systems may join an interoperating system. Ethereum's layer 2 networks~\cite{stark2018making} can be considered a set of interoperable chains. However, chains outside the Ethereum network cannot natively communicate with these layer 2 chains, thus limiting the scope of interoperability. The solution must account for the possibility of unforeseen networks joining the system to offer global interoperability. 

The National Institute of Standards and Technology (NIST) defines \emph{blockchain interoperability} as: ``a composition of distinguishable blockchain systems, each representing a unique distributed data ledger, where atomic transaction execution may span multiple heterogeneous blockchain systems, and where data recorded in one blockchain are reachable, verifiable, and referable by another foreign transaction in a semantically compatible manner"~\cite{yaga_blockchain_2018}. 
In our definition of blockchain interoperability, we consider the various features essential for a blockchain to interoperate with another blockchain, or another legacy solution, especially when taken across multiple application domains. Therefore, we define \textbf{blockchain interoperability} as the ability for a given blockchain to join multiple blockchain networks and other software systems in a \textit{scalable} manner that extends the NIST definition. Essentially, a blockchain interoperability solution must enable a blockchain to conduct transactions utilizing information housed in another information system in a verifiable manner.
The above definition encompasses common interoperability use cases, such as a token exchange and many other decentralized applications. It also contains uses such as reading stock market data from a non-decentralized system in a verifiable way. We consider this definition because it is the broadest definition that retains the properties of a host blockchain and enables its extensibility.

\subsection{Blockchain Interoperability Use Cases}

As blockchain technology evolves, there is a constant stream of new projects with new promises. However, the current siloed nature of blockchain systems may interfere with the projects' ability to scale. Blockchain interoperability can help isolated ecosystems trust and work with each other to maximize their utility. The types of information available in a blockchain network can be classified into two broad categories: assets (cryptocurrencies, tokens) and data. This section examines how these two domains can benefit from blockchain interoperability.

\subsubsection{Asset Movement}
The ability to mine cryptocurrencies and trade digital assets that are not regulated by a central authority constitutes the main reason for the popularity of blockchains. However, with an increasing number of projects in the ecosystem, the ability to transfer these siloed assets between different networks will become ever more critical.
Pillai et al.~\cite{pillai2022cross} define three types of use cases for assets in a cross-chain environment. These are (1) asset exchange, (2) asset transfer, and (3) asset migration. Asset exchange is a two-way operation in which entities on different blockchains can exchange an agreed-upon amount of their respective assets. Asset transfer is an operation in which one entity moves its asset from one blockchain to another. Asset migration is when an entity \textit{permanently} moves its asset between two blockchains. Although the last two use cases sound similar, they are, in fact, different in the protocols used for each case. Asset movement across blockchains can offer advantages that a single blockchain cannot--for example, when an application developer wishes to develop a decentralized application (DApp) without making its users bear the burden of the high network fee of Ethereum. An interoperability solution can allow the developer and the potential users to seamlessly convert their assets onto a different network with lower gas fees. The developer can then attract a more extensive user base while the users benefit from a more affordable service. 

\subsubsection{Web3 Applications}
Web3 provides an application substrate for a new generation of Web applications, bringing decentralization and tokenomics to the current Web. Instead of traditional databases or servers, blockchain(s) can act as the database for the Web3 applications and program a set of constructs for the applications to interface with it. A significant obstacle to achieving a fully connected Web that lets users read, write and own assets using blockchains is that most blockchains are siloed ecosystems. A well-thought-out interoperability solution must be in place for an application to seamlessly use multiple blockchains. For example, one could build a DApp for selling customized shirts, and this application could connect different vendors to the end user to deliver the final product. If every vendor, shipper, shirt manufacturer, shirt printer, etc., has a different blockchain system running for its operations, such an application would need to be able to communicate with every party through their respective blockchain. If there is an interoperability solution between blockchains, the application will utilize it to deliver the final product to the user. This example is simple, but it shows a glimpse into the possible future of an Internet economy powered by decentralized solutions. While blockchains possess several appealing properties, their gamut of possibilities is limited without interoperability since all stakeholders cannot be expected to use the same blockchain platform in a truly decentralized environment. 

\subsubsection{Secure and Accountable Information Exchange}

Handling sensitive information in an accountable manner is another critical task that can be driven by blockchain technology, which is made possible by two aspects of blockchain systems. 
First is unique identifiability across the entire network: in the case of blockchain, the interconnected network. Each device has a unique `name' in the network, allowing applications such as search engines or routing protocols to optimize their operations. For example, this can be the assignment of a unique identifier to each wallet in the specific network. Identifiability inside each network is a part of the blockchain mechanism. However, sharing this identity information outside each blockchain's silo is an unsolved problem. If a unique identity across the entire blockchain space is established, it can lead to more efficient data routing and more security for cross-chain operations.

Second is the ability to verify the off-chain identities of individuals in a permissioned manner. Know Your Customer (KYC), and Anti Money Laundering (AML) are essential properties in traditional finance to prevent crime and misconduct, primarily when the trade partners are not known or are untrusted. These properties are more difficult to enforce in the blockchain space due to the pseudo-anonymous nature of the underlying technology. Some centralized exchanges, such as Coinbase, 
require KYC for customers to create an account. However, in the space of Decentralized Exchanges (DEX), KYC is not a requirement. Managing such sensitive information in a decentralized environment is a tricky task. Chainlink~\cite{breidenbach2021chainlink} is an example of a blockchain solution that addresses this problem by providing verified external data through its network. Some enterprises are also evaluating permissioned blockchains to manage such sensitive data. However, for a blockchain to use the data provided by such solutions, it must have some degree of interoperability, which can be addressed by establishing communication between oracle providers and the consumer blockchain. Furthermore, permissioned sharing of information can be enabled if the chains can verify the identity -- thus trustworthiness -- of the entities across different blockchains. The problem of identity is not limited to financial services. Other industries, such as healthcare or logistics and supply chains, can also benefit from sharing this type of sensitive data, potentially leading to innovations that were, heretofore, unavailable.

\subsection{Our Contribution}

This work provides an overview of the current landscape of blockchain interoperability by categorizing nine existing projects. While past research has focused on a theoretical categorization of possible interoperability approaches and pointing out some real-life examples, we approach the problem with a bottom-up approach by examining the existing solutions and analyzing and categorizing them. We use and expand upon the definitions provided in previous literature and apply them to the existing interoperability solutions to understand their underlying enabling mechanisms better.


\section{Previous Work}

Siris et al.~\cite{siris2019interledger} presented a literature survey of interoperability approaches. The authors analyze techniques such as cross-chain transactions, bridges, sidechains, and protocols like the W3C Interledger Protocol (ILP)~\cite{hope2016interledger} and classify them by their supported assets, scalability, trust, and cost. Kannengiesser et al.~\cite{kannengiesser2020bridges} characterized approaches toward blockchain interoperability and their strengths and weaknesses. The authors collected literature related to cross-chain transactions, extracted common design patterns used in the literature, and assessed the strengths and weaknesses of their administration, flexibility, performance, and security properties.

Belchior et al.~\cite{belchior2021survey} conducted a systematic review of blockchain interoperability by collecting and evaluating the related literature. The authors reviewed the existing mechanisms by classifying them by use case, payload, and degree of decentralization of each framework. Furthermore, they outlined the case for blockchain interoperability research and the need for adaptability driving the growth of both new blockchains and blockchain interoperability. They also explored various definitions and constructed a concept map for blockchain interoperability. Overall, their work highlighted the fragmented nature of blockchain interoperability research and the need for further classification of solutions. The authors highlighted several risks for interoperability solutions, namely, what if (1) security vulnerabilities are uncovered?, (2) the chosen blockchain does not support new use cases?, (3) the rest of the industry adopted an improved version of the technology?, or (4) an organization’s users exist on another solution? Mitigating such risks requires adaptability by either improving the existing blockchain to make it adequate for transacting with other blockchains or integrating it with an existing blockchain, thus making the blockchain extensible.

Lohachab et al.~\cite{lohachab_towards_2022} explore the generations of blockchain technology and introduce the need for blockchain interoperability. They describe potential interoperability architectures and taxonomy to compare the state-of-the-art interoperability research, concluding with the challenges for future research directions to achieve blockchain interoperability. Pillai et al.~\cite{pillai2022cross} characterize the different layers of a blockchain interoperability solution, providing a guideline for building blockchain interoperability solutions depending on the use case, such as data relay or a crypto exchange, along with the different mechanisms that can be used to achieve these goals. Twelve properties that should be considered when evaluating a blockchain interoperability solution are described by Koens et al.~\cite{koens2019assessing} in analyzing two existing solutions, Polkadot~\cite{polkadot} and Cosmos~\cite{cosmos}. Based on these metrics, the authors outline the similarities and differences in terms of the 12 properties. Finally, Singh et al.~\cite{singh2020sidechain} focus on sidechain mechanisms to achieve interoperability. They outline the different sidechain designs, their use cases, and their strengths. 




\section{Layers of Blockchain Interoperability}

The blockchain ecosystem is made complete by the different layers. The fundamental building blocks of a blockchain are classified as the consensus, networking, data, and hardware layers. With the introduction of Decentralized Applications (DApps), applications can now consume information from the consensus layer of blockchains. In addition to these layers, modern blockchain developments have introduced another abstraction layer. This layer handles the operations between the application and consensus layers, allowing the data to flow between different blockchains for a particular application in a blockchain-agnostic manner. Interoperability between blockchains can happen on separate layers. This work examines applications that either facilitate data flow between blockchains through the messaging layer or applications that provide the tools for building new blockchains. 











\subsection{Hardware Layer}
Also known as the infrastructure layer, this layer refers to the hardware that runs the node client software for the blockchain. 
This node's security depends on the blockchain's permission level. In a public network, anyone may maintain a node. However, in a permissioned network, only trusted entities are allowed to run a node.

\subsection{Transaction Layer}
This layer refers to the transactions and blocks in a blockchain. A block is formed by grouping together different transactions, and it must be verified by a fraction of nodes greater than a certain threshold to be accepted into the ledger. In this layer, the nodes listen for transaction information and batch them together into a block. 
The transaction layer is where differences in most blockchain projects start to appear. Different protocols specify the ordering mechanism of transactions and the size of each block, preventing them from being interoperable. The encryption and hashing of data also happen in this layer. Each blockchain can use a different algorithm to encrypt its payload, resulting in varying degrees of difficulty and security guarantees. This inconsistency adds to the interoperability challenge. 

\subsection{Networking Layer}
The networking layer consists of node clients that use peer-to-peer networking to finalize the input to the ledger. It involves nodes submitting their work to each other and checking it, where the data protocol may vary for each network, thus requiring different fields or routing protocols. 

\subsection{Consensus Layer}
The consensus layer is where the nodes check each other's work and agree on a result to add to the ledger. Blockchains can have different consensus mechanisms, such as Bitcoin's \textit{Proof of Work} (PoW), Ethereum's \textit{Proof of Stake} (PoS), or Solana's \textit{Proof of History} (PoH). Even if some blockchains share the mechanism, the protocol can differ by their validation threshold. The differences in this layer can also impact the security threshold of individual blockchains, making it difficult to achieve safe interoperability.

\subsection{Messaging Layer}
In a traditional blockchain system, the messaging layer would be a part of the application layer. However, as the need for blockchain interoperability grows, some projects, such as Overledger~\cite{verdian2018quant} or Hyperledger Cactus~\cite{hyperledger-cactus}, have added this layer. Such an abstraction layer, which is often located between the consensus and application layer, can be an entry point between multiple blockchains working with a single multi-DApp. Therefore, a solution must choose a standard protocol for sharing consensus results between numerous blockchains. Overledger, for example, solves this problem by writing the message payload to the `comment' field of different blockchains.

\subsection{Application Layer}
This layer interfaces with the end users of DApps, and delivers the information passed from the blockchain to its DApps. It can also involve application-supporting features, such as smart contracts.

\begin{figure}[!htbp]
    \centering
    \includegraphics[width=.8\linewidth]{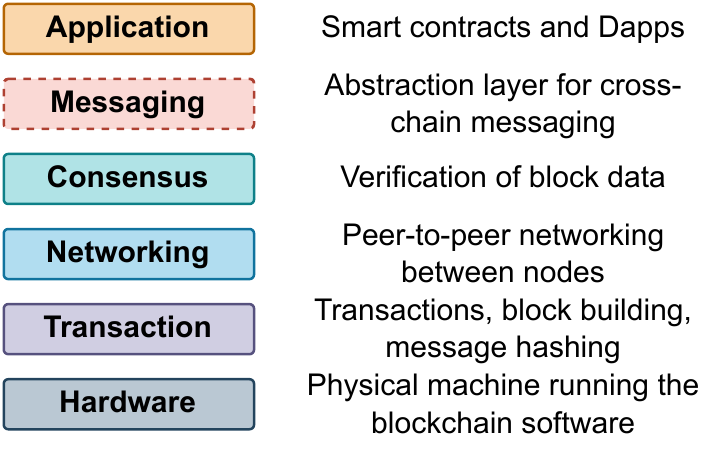}
    \caption{Blockchain System Layers}
    \label{fig:blockchain_layers}
\end{figure}

\section{Variations in Blockchain Implementations}
\begin{figure}[!htbp]
    \centering
    \begin{subfigure}[t]{.48\linewidth}
      \includegraphics[width=\textwidth]{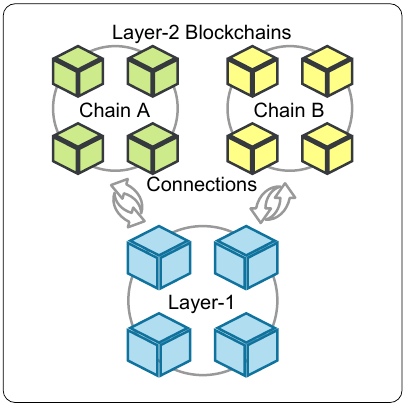}
      \caption{Layer 2}
      \label{fig:vari_layer2}
    \end{subfigure}
    \hfill
    \begin{subfigure}[t]{.48\linewidth}
      \includegraphics[width=\textwidth]{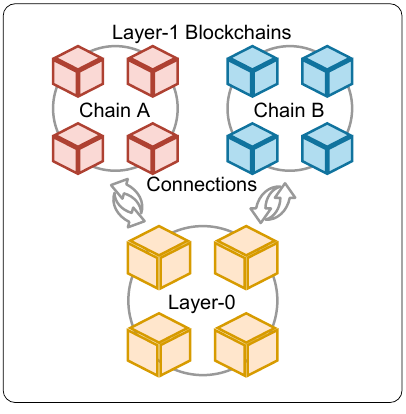}
      \caption{Layer 0}
      \label{fig:vari_layer0}
    \end{subfigure}
    \begin{subfigure}[t]{\linewidth}
      \includegraphics[width=\textwidth]{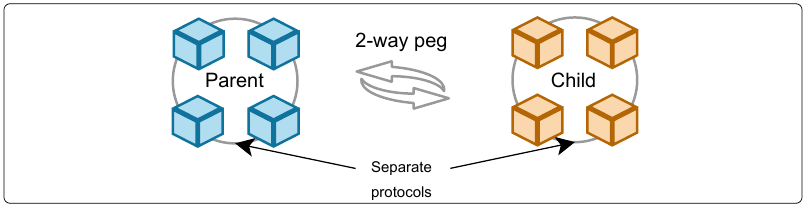}
      \caption{Sidechain}
      \label{fig:vari_sidechain}
    \end{subfigure}
    \hfill
 \end{figure}

Computational efficiency and interoperability are key issues addressed by these new technologies. As the blockchain space becomes increasingly complex, it is necessary to separate layers within blockchains based on their roles. In our survey of interoperability solutions, there are three types of mechanisms, which are \textit{sidechains}, \textit{layer-2 blockchains}, and \textit{layer-0 blockchains}. A simplified diagrams of these categories are shown in Figures~\ref{fig:vari_layer2}, \ref{fig:vari_layer0} and~\ref{fig:vari_sidechain}.

\subsection{Sidechains}
Sidechains, one of the earliest approaches towards interoperability, were first introduced by Back et al.~\cite{back2014enabling}. This mechanism uses the existing blockchain network (the parent chain) and allows assets to move between various networks. A sidechain is a separate blockchain and can be connected to the parent chain to allow asset movement. A two-way peg mechanism~\cite{two-way-peg} makes the asset movement possible, which enables the chain-specific asset to be \textit{wrapped} into the supported asset on the destination chain. This mechanism is characterized by the fact that they are connected to their parent chains and run parallel to them. 

\subsection{Layer-2 Blockchains}
Layer-2 blockchain is another form of blockchain that runs on top of a parent (or layer-1) blockchain. It shares many similarities with the sidechain mechanism in that it runs parallel to its parent chain and is connected to it. The difference is that while sidechains may maintain their state and security protocol separately, layer-2 blockchains rely on the state of the layer-1 blockchain. The main goal of layer-2 blockchains is to extend the functionality of layer-1, as well as enable faster transaction speed by relying on layer-1's infrastructures. Layer-2 can be implemented using mechanisms other than a two-way peg~\cite{two-way-peg}, such as rollups~\cite{rollups}, or bridges~\cite{kannengiesser2020bridges}.
Sometimes the terms \textit{sidechain} and \textit{layer-2} are used interchangeably in the literature. Our interoperability landscape characterization uses the above definitions (dependence on the layer-1 chain) to avoid confusion between these two mechanisms.

\subsection{Layer-0 Blockchains}
Some interoperability projects, such as Polkadot and Cosmos, use this term to describe their implementations. This idea is an extension of the relationship between layer-1 and layer-2 blockchains. Similar to how the layer-1 network supports the layer-2 network by providing consensus and security, layer-0 chains provide functionalities that the layer-1 chain can use. Thus, layer-0 blockchains are \textit{meta-chains} providing infrastructure and interoperability between the depending layer-1 chains.

\begin{table}[!htbp]
\tiny
\resizebox{\linewidth}{!}{%
\begin{tabular}{|cc|c|c|c|}
\hline
\multicolumn{2}{|c|}{\textbf{Connection Method}} & \textbf{Payload} & \textbf{Network Permission} & \textbf{Projects} \\ \hline
\multicolumn{1}{|c|}{\multirow{3}{*}{On Chain}} & layer-0 & Crypto, Data & All & Polkadot, Cosmos, \\ \cline{2-5} 
\multicolumn{1}{|c|}{} & \multirow{2}{*}{layer-2} & Crypto & All & Loom \\ \cline{3-5} 
\multicolumn{1}{|c|}{} &  & Crypto & Public & Hybrix \\ \hline
\multicolumn{1}{|c|}{\multirow{3}{*}{\textit{Off-Chain}}} & \multirow{2}{*}{Gateway} & Crypto & Public & Poly Network \\ \cline{3-5} 
\multicolumn{1}{|c|}{} &  & Data & All & Overledger, Cactus, Firefly \\ \cline{2-5} 
\multicolumn{1}{|c|}{} & Oracle & Data & All & Chainlink \\ \hline
\end{tabular}%
}
\caption{Blockchain Interoperability Project Categorization.\\The connection method contains the on-chain and off-chain characteristics and further breaks that down to how it is achieved (layer-0, layer-2, gateway, or oracle).
The payload column refers to the type of data the framework handles, which includes crypto-assets (\textit{crypto}) and arbitrary data (\textit{data}). The network permission column describes the permission level the framework can handle, such as \textit{public} or \textit{private} or \textit{all} (the framework supports both types of permission).}
\label{tab:taxonomy_table}
\end{table}

\section{Characterization of Existing Blockchain Interoperability Approaches}
\label{sec:existing_apps}

Blockchain interoperability is a relatively new field, and advances are happening rapidly. There is a variety of solutions with different approaches for achieving interoperability. We analyze a range of blockchain interoperability projects and present an analysis of nine projects with exciting features introduced in \Cref{sec:solutions}.
We then compare the differences in these approaches, namely, what they promise, how their goals are achieved, how they can be extended in the future, and their strengths and weaknesses. Table~\ref{tab:final_table} provides an overall summary of our analysis.

\subsection{Summary of Solutions}
\label{sec:solutions}
Blockchain interoperability presents many new paradigms, ranging from simply transferring assets to being able to share sensitive data in an accountable manner over the Internet. Although there are some overlaps in the goals of each framework discussed, each framework incorporates a unique goal and a method of achieving it.

\smallskip
\noindent\textbf{Overledger~\cite{verdian2018quant}:}
Overledger's official purpose is to provide a `messaging layer' for blockchain-based applications, which is agnostic to each blockchain's consensus mechanism. Instead, it allows the different chains to send messages to each other through a standardized format. Quant Network, the company behind Overledger, describes the project as an \textit{``operating system of blockchains''}, as it can serve as the entry point for existing payment systems and blockchains to be abstracted for the application layer. 

\smallskip
\noindent\textbf{Hyperledger Cactus~\cite{hyperledger-cactus}:}
Hyperledger Cactus is an enterprise-grade application that allows for seamless cross-chain transactions. On top of achieving interoperability for assets, it also looks into the interoperability of data between blockchains, such as sharing inventory data between permissioned enterprise networks. 

\smallskip
\noindent\textbf{Hyperledger Firefly~\cite{hyperledger-firefly}:}
Hyperledger Firefly is an API abstraction layer for applications based on the blockchain, similar to Overledger. It aims to allow for permissioned data passing among different chains. Compared to Hyperledger Cactus, this project focuses more on the application development side, as it seeks to provide a software development kit for building DApps based on multiple chains.

\smallskip
\noindent\textbf{Polkadot~\cite{polkadot}:}
Polkadot is a layer-0 solution for blockchain interoperability. It enables the communication between different blockchains through what is called ``parachains'' and the ``relay chains''~\cite{schulte2019towards}. It looks to address scalability in computing and networking while working with systems with diverse needs and guarantees in the blockchain space.

\smallskip
\noindent\textbf{Cosmos~\cite{cosmos}:}
Cosmos shares a similar goal as Polkadot, providing a layer-0 solution for interoperability. It aims to address the drawbacks of many current blockchain networks, such as energy inefficiency and low performance, and bring all networks together through a parallelized structure called ``zones.''

\smallskip
\noindent\textbf{Loom~\cite{loom_network}:}
The Loom Network is an Ethereum sidechain powered by the native \texttt{LOOM} token, a native ERC-20 token. It strives to achieve interoperability by connecting different blockchains to the `basechain' using a gateway mechanism. Loom tokens can be burned and minted on the source and destination chain using these gateways, allowing liquidity flow across various blockchains.

\smallskip
\noindent\textbf{Hybrix~\cite{hybrix}:}
Hybrix is a peer-to-peer (P2P) protocol to achieve interoperability between blockchains. A node on a Hybrix system runs software called \texttt{hybrixd}, which includes connection mechanisms for supported blockchains, a data abstraction mechanism, and a secure signing mechanism to enable a connection between interoperating blockchains.

\smallskip
\noindent\textbf{Chainlink~\cite{breidenbach2021chainlink}:}
Chainlink is a network of \textit{Decentralized Oracle Networks} (DONs). It aims to provide reliable data -- both from on-chain and off-chain -- for other blockchains to consume. With the announcement of Chainlink v2 \cite{breidenbach2021chainlink}, the project has introduced a new term called \textit{hybrid smart contracts}, which is an abstraction of on-chain computation using off-chain data from DONs. 

\smallskip
\noindent\textbf{Poly Network~\cite{poly_network}:}
Finally, Poly Network is another project that follows the hub and spoke architecture (\Cref{fig:int_mode_hub}) to provide interoperability. Poly Network itself is not a blockchain, but it provides the protocol for nodes of blockchains participating in Poly Network to create a consortium blockchain that has shared storage. 

\subsection{Integration Mode}
\label{subsec:integration_mode}

\begin{figure}[!htbp]
    \centering
    \begin{subfigure}[t]{.49\linewidth}
        \centering
        \includegraphics[width=\textwidth]{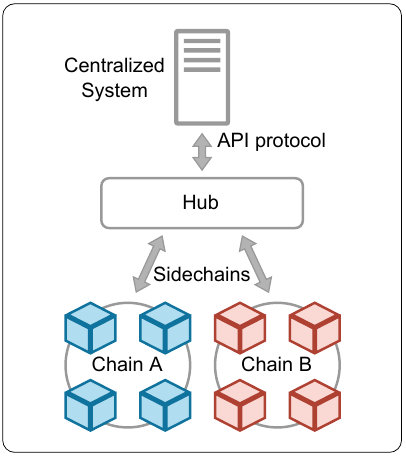}
        \caption{Hub and Spoke}
        \label{fig:int_mode_hub}
    \end{subfigure}
    \begin{subfigure}[t]{.49\linewidth}
        \centering
        \includegraphics[width=\textwidth]{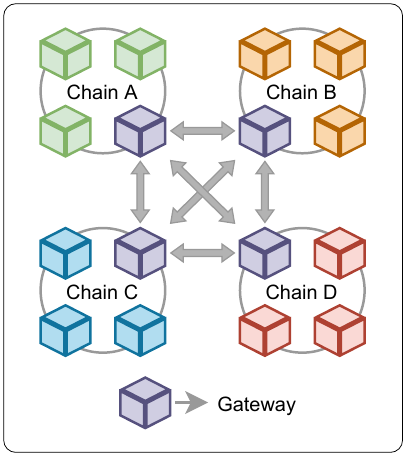}
        \caption{Decentralized Relay}
        \label{fig:int_mode_relay}
    \end{subfigure}

    \begin{subfigure}[t]{\linewidth}
        \centering
        \includegraphics[width=\textwidth]{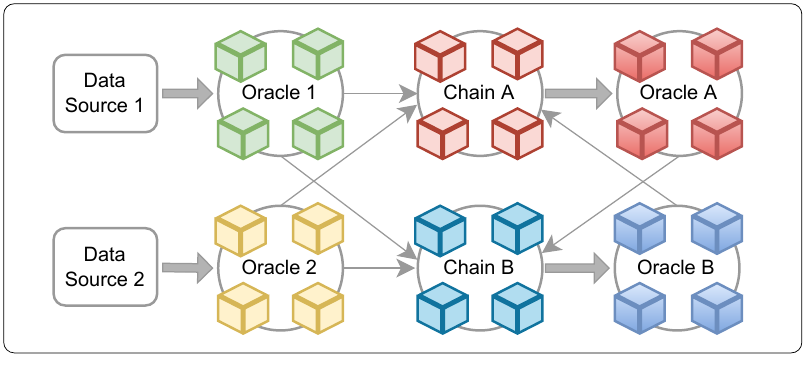}
        \caption{Decentralized Oracle}
        \label{fig:int_mode_oracle}
    \end{subfigure}
 \end{figure}

Pillai et al.~\cite{pillai2022cross} introduced the term ``cross-blockchain'' to define the method through which the blockchains are connected. This property is essential because it affects many other properties of an interoperability solution. If a solution introduces a new blockchain to solve the problem, it may be able to guarantee more robust security or governance. However, it presents a new issue of scaling the said blockchain to the ever-evolving blockchain space. If a solution uses an off-chain mechanism to connect, it may have a lower security guarantee, but it can also benefit from a reduced scaling overhead. In general, we see three different methods of approaching this problem, namely \textit{Hub and Spoke}, \textit{Decentralized Relays}, and \textit{Decentralized Oracles}.


\medskip
\noindent\textbf {Hub and Spoke:}
The hub and spoke approach is a commonly used architecture for constructing large networks. Figure \ref{fig:int_mode_hub} shows a simple architecture diagram. This approach connects individual `nodes' -- blockchains in this case -- to a central hub through which they can interoperate. Some projects implementing this architecture are Polkadot, Cosmos, Loom, and Poly Network. 


In Polkadot, the \textit{parachains} are analogous to the spokes and the \textit{relay chain} to the hub. For Cosmos, the \textit{Zones} is equivalent to the spokes and the \textit{Hubs} to the hub. 
Poly Network also uses this approach to enable the interoperability of cryptocurrencies of connected networks. The \textit{relayer}, as it is known in Poly Network, is in charge of converting the currency and syncing with the main network. Loom implements this design by using an Ethereum sidechain and allowing its tokens to be interoperable in other supported sidechains.


\medskip
\noindent\textbf{Decentralized Relay:}
The decentralized relay architecture is a design in which networks can communicate through `gateways.' 
A blockchain gateway is an entry point to the underlying ledger with read/write access. In this design, each gateway is assigned a particular blockchain to connect to the main network or handle connections between the main network and all other blockchains. Figure \ref{fig:int_mode_relay} shows a simple diagram describing this architecture. For projects such as Overledger, Hyperledger Cactus, and Firefly, the software must run on a trusted machine -- not necessarily an active network node -- and use connectors for each blockchain. Hybrix also uses this architecture to relay messages from one blockchain to another through a P2P gateway mechanism.
At a glance, this architecture can appear similar to the Hub and Spoke architecture, as it achieves a similar network scheme of connecting different chains through a designated entry point. To avoid confusion, we differentiate these by their usage of additional blockchains. Then, the Hub and Spoke approach is an architecture in which both the hub and spoke are different blockchains that sync with the target chains. On the other hand, a decentralized relay is a piece of software that connects blockchains \textit{without} introducing additional blockchains, meaning that it does not necessarily have to deal with the state of the networks as long as it forwards data between blockchains.

\medskip
\noindent\textbf{Decentralized Oracles:}
An oracle in the blockchain space refers to an entity that allows a blockchain to access out-of-chain information. This information can be both \textit{on-chain} from another blockchain, such as the amount of liquidity present in the network, or \textit{off-chain}, such as foreign exchange rates. 
If a blockchain depends on information from a centralized oracle service, the validity of the entire blockchain ledger can be compromised by compromising the oracle's data source. Decentralized oracles try to tackle this problem by adding redundancy at the oracle level by having multiple nodes serve the same data only if it can be agreed upon by the validators and verified.
Previous works in analyzing blockchain interoperability have classified the Oracle approach as a \textit{mechanism} used to achieve interoperability, similar to that of other mechanisms such as Hashed Time Lock Contracts~\cite{dcunha2021blockchain} or Burn-and-Mint \cite{belchior2021survey}. However, unlike those mechanisms, oracles are not necessarily crypto-centric. It can convey information related to transactions, but it can also communicate any other kind of data. In this work, we view decentralized oracles as an interoperability solution because of the wide range of data they can transfer between blockchains and the traditional world. Chainlink~\cite{breidenbach2021chainlink} implements this approach by connecting various Decentralized Oracle Networks (DONs) to create a meta-chain of oracles that can provide data feeds to its target blockchains.

\subsection{Validation}
Blockchains can have different security assumptions that need to be validated. 
Existing solutions take different approaches towards this problem:


\medskip
\noindent\textbf{Native Validation:}
Native validation is defined as validation happening inside the interoperability solution. This approach applies to solutions implemented in their chains, which can have their state validated. Several projects, such as Polkadot, Cosmos, Loom, Chainlink, and Poly Network, use this approach.

\smallskip
\noindent\textit{Polkadot:} Polkadot's relay chain state is maintained by its validator nodes. To prevent malicious behavior from validators, Polkadot has a unique role called \textit{fishermen} who are in charge of checking the blocks signed by the validators. If it turns out that the fishermen did indeed detect misbehavior, they are rewarded by the network. The reward amount scales with the number of validators who signed off on the illegal block, as it implies higher risk in the network \cite{polkadot}. In each parachain, collators are the nodes responsible for working with the validators. As the name suggests, the collators combine the transactions by building a block for the validators.

\smallskip
\noindent\textit{Cosmos:}
In Cosmos, each \textit{zone} updates its parent \textit{hub} with the state of its blockchain. The zones also update their status based on the hub, allowing for bi-directional information flow \cite{cosmos}. When a zone needs to send or receive information from another zone, it posts a Merkle tree proof of the sender on the hub, which the destination zone receives. Similarly, the destination zone also publishes evidence that the information was received. This mechanism is similar to how sidechains communicate by establishing a bi-directional channel for proof-of-existence transactions. 

\smallskip
\noindent\textit{Loom:}
The Loom base chain uses Delegated Proof of Stake (DPoS), an improvement over the commonly used Proof of Stake (PoS) mechanism, to validate the data coming from outside. Unlike PoS, in which the validator is chosen with randomness proportional to each user's staked amount, DPoS allows users to \textit{vote} for new validators, each vote worth the amount they have staked. 

\smallskip
\noindent\textit{Chainlink:}
As an oracle framework, Chainlink must verify that it serves the correct data, whether off-chain or on-chain. Before the data is relayed to the requesting blockchain, validation is done by selecting random nodes of each DON to compare the data against its own.

\smallskip
\noindent\textit{Poly Network:}
Validation happens in Poly Network by recording the block header of the relayed data. Using the Merkle root inside the block's header, the Poly Chain (the main chain) re-verifies the block's validity. Once this block is verified and recorded in the relay chain, the data can be sent to other chains, completing the cross-chain operation. 

\medskip
\noindent\textbf{Third-party Validation:}
Third-party validation refers to a validation mechanism in which the task of validating is delegated to a third party who is not a part of the interoperability solution. A federation of validating nodes is commonly used to provide a robust security guarantee. While convenient because different validators can be switched in and out, this approach can also introduce a security bottleneck if it needs to be decentralized enough compared to the connected blockchains.
Projects that use this approach are Hyperledger Cactus and Firefly, both of which make use of \textit{ledger plugins} (called \textit{blockchain plugins} in Firefly). The choice of validator service is left up to the user, meaning that the security of these projects' instances depends on the validator service.

\medskip
\noindent\textbf{Hybrid Validation:}
Hybrid validation is when verification happens through a different mechanism than extending the connected chains' state. This mechanism is applicable when the data being passed in the interoperability protocol are not \textit{assets} but pieces of arbitrary \textit{data}. A typical implementation of this mechanism uses the \textit{message} field in transaction data. By setting a protocol for reading and writing to this field, the interoperability solution can ensure that the data it receives is original. Two projects that use this approach are Overledger and Hybrix. 
Overledger uses a proprietary algorithm known as \textit{TrustTag} to generate and verify hashes for messages. When a transaction needs to be sent off-chain, the user's identity and transaction details are encrypted using the application's public key and added to the metadata field. The digest is verified using the application's private key when this message arrives at the application layer. This mechanism allows the app to confirm that the message has not been tampered with and provides the ability to assign unique IDs for each user. This technology is closed source and is patented by Quant, the company behind Overledger.
%
Like Overledger's message verification mechanism, Hybrix also uses the message header field to validate the originality of the incoming data. To achieve this, empty transactions that carry data are composed and pushed to each blockchain. The first block contains a \textit{genesis recipe}, which establishes the interoperability protocol. The cross-chain transactions sent after this genesis recipe is checked using its hash value.

Poly Network allows each chains to validate each other by checking the transaction's block header in the source chain. This is made possible by the Poly Chain's synchronized header mechanism which synchronizes with every participating chain's block height and records the information.




\subsection{Upgradability}
Upgradability measures how the framework can stay flexible to keep up with the evolving technologies in the blockchain space. A good step of upgradeability is that the framework should continuously upgrade itself to keep up with the evolving technologies and provide a reliable solution. One good example of a measure for this property is the possibility of updates causing a hard fork. For instance, if the community cannot agree on the proposal leading to a hard fork, it can cause a divide. Such an event can cause the interoperability solution to have fewer interoperating blockchains and cause further fragmentation in the blockchain space. This property is also closely related to the security of the framework, as there can be unexpected bugs in the software. 
Upgradability in interoperability solutions can be broadly categorized into two groups: Software-based and Blockchain-based.





\medskip
\noindent\textbf{Software-based:} This category applies to solutions that provide the software to construct or connect to the network, such as Overledger, Cactus, Firefly, and Hybrix. These solutions implement a microservice-based architecture, meaning that the \textit{connector} service for each network can be upgraded separately without impacting the other functionalities.

\medskip
\noindent\textbf{Blockchain-based:} This category applies to solutions that leverage a new blockchain to achieve interoperability, including projects such as Polkadot, Cosmos, Loom, Chainlink, and Poly Network. Upgradability is an essential feature in this category of solutions because they can be susceptible to a hard fork if the community does not unanimously agree with a substantial update. Polkadot addresses this problem with its protocol. Its governance is a meta-protocol, meaning that the governance can define upgrades without requiring a hard fork. 
Chainlink addresses this problem by providing a service called \textit{contract-update} to account for unforeseen changes. This service allows users to modify the existing smart contract, publish a new one, and update the DON connections to point to the revised contract. Cosmos, Loom, and Poly Network do not specify plans for unforeseen changes to the network. 
Thus, with the exceptions of Polkadot and Chainlink, blockchain-based solutions may face a hard fork when there are essential requirements to upgrade.

\subsection{Adoptability}
Adoptability is how new networks can be adopted into the existing framework, allowing it to scale further in the blockchain space.
The distinction between upgradability and adoptability can be characterized as follows. Upgradability refers to the innate ability of the framework, while adopotibility refers to how external networks can join the framework. There are many variables to consider when considering a framework's adoptability, ranging from the data protocol to the network structure. An interoperability framework with higher adoptability is more likely to gain a larger community, which is crucial in the current blockchain environment. 


\medskip
\noindent\textbf{Hub and Spoke:} Projects that implemented the hub and spoke architecture can have more complicated requirements for new networks to join the interoperability network. These requirements arise from the hub's blockchain need to stay secure. Depending on the implementation of the main chain's validation and consensus, each project has a slightly different requirement for new blockchains.

\noindent\textit{Polkadot:} Anyone \textit{can} build a parachain for Polkadot \cite{polkadot}. For a parachain to become a part of the Polkadot ecosystem, it must have validators. To maintain the integrity of the main network when adding new networks, Polkadot requires that the existing stakeholders of the network be incentivized by some means -- whether that be paying a fee or providing a new feature. For connecting a new chain that does not conform to Polkadot's governance, such as Bitcoin, Polkadot provides a ``bridge chain.'' Additionally, Polkadot runs a test network known as ``Kusama'' to provide a sandbox for developers to test their blockchains.

\smallskip
\noindent\textit{Cosmos:} 
Anyone can build a zone for Cosmos. Cosmos separates the governance between the hub and zones, allowing its zones' failures to be isolated from the entire network. Compared to Polkadot, Cosmos has a lower entry condition as the zone creator can choose their validators for the zone instead of being designated validators by the main network. 

\smallskip
\noindent\textit{Loom:} Loom network's main chain exists as a layer-2 Ethereum blockchain and is supported by the Loom ERC-20 token. Loom has achieved interoperability by allowing different `versions' of this token to be transferred between various blockchains using smart chain mechanisms such as burning and minting. Although Loom makes no specific requirements for a blockchain to be supported, it can be assumed that the new blockchain must support some form of ERC-20 compatible tokens and some form of smart contracts.

\smallskip
\noindent\textit{Poly Network:} The relayer function needs to be implemented for a new blockchain to join the Poly Network environment. The new network must also support light client verification mechanisms such as Simple Payment Verification (SPV) or Light Ethereum Subprotocol (LES) to implement the validation step in the relay chain. 

\medskip
\noindent\textbf{Decentralized Relays:} Decentralized relays such as Overledger, Hyperledger Cactus, Hyperledger Firefly, and Hybrix use the comment field to facilitate interoperability. Thus, a critical requirement for a new network to join these systems is that it must support a comment field in the transaction data. For Hybrix, the new blockchain must also be public for validation purposes.

\smallskip



\smallskip


\smallskip

\medskip
\noindent\textbf{Decentralized Oracles:} Chainlink does not have any requirements because its role is simply to serve the data. However, some security requirements exist for new DONs to join its meta-network \cite{breidenbach2021chainlink}.


\subsection{Governance}
The governance of a decentralized system is crucial in ensuring a bright future for the system by making the best decisions. When putting together different networks, this can become an interesting problem. Should the sub-networks (lower-level blockchains) have a say in the governance of the overall interoperability framework? Or should this framework have a separate community that makes the decisions? We show two significant ways of dealing with this problem. Overledger, Cactus, Firefly, and Hybrix take the off-chain governance approach, in which connected blockchains' governance does not affect the framework. On the other hand, Polkadot, Cosmos, Loom, Chainlink, and Poly Network implement a governance protocol made up of the validators in the system. 

\medskip
\noindent\textbf{On-chain:} On-chain governance is applicable for solutions that maintain the main blockchain network that other networks connect through. 
\smallskip

\noindent\textit{Polkadot:} Polkadot governance is composed of a `council' and a `technical committee.' A council is a group of elected stakeholders, of which there are currently 13 members. The council is responsible for: 1. managing treasury, 2. proposing referenda, 3. canceling dangerous referenda, and finally, 4. voting for the technical committee members. Only teams or individuals who have added successful implementation to Polkadot can be a part of the technical committee. The technical committee can submit emergency proposals and veto a proposal. 

\smallskip
\noindent\textit{Cosmos:} Cosmos' governance is responsible for managing updates to the network, changing protocol parameters, and using community pool funds. Each validator needs to deposit a minimum amount of ATOM tokens. A validator may be deactivated for a certain amount of time if a validator does not vote. On top of the Hub's governance, each zone may also have its governance. Nevertheless, the Hub's governance will reign superior, acting as a constitution for zones inside the Cosmos ecosystem. 

\smallskip
\noindent\textit{Loom:} Loom Network's governance is based on the native LOOM token. Validators and delegators stake the tokens to secure the network. Application developers who use Loom's base chain also pay for their service in LOOM, even though the individual users do not need to pay. The token can also be bonded when a developer adds a new chain to the network to discourage malicious behavior.

\smallskip
\noindent\textit{Chainlink:} With the announcement for v2 of Chainlink, no specific governance protocol has been announced for the network. However, the governance model will be \textit{evolutionary} \cite{breidenbach2021chainlink}, meaning that the governance mechanism itself can be subject to change. Each DON may also have its governance mechanism to account for DON-specific changes. 

\smallskip

\noindent\textit{Poly Network:} Governance happens on its main chain, called the Poly Chain \cite{poly_network}. However, Poly Network does not issue a token, so it remains to be seen how the network members can partake in the governance of the main chain.

\smallskip
\noindent\textbf{Off-chain:} Interoperability solutions that provide the software rather than another blockchain do not have on-chain governance. These solutions are Hyperledger Cactus, Hyperledger Firefly, and Hybrix. They are designed to be used by a consortium of entities to run their nodes and form an off-chain network. Overledger is an exception because it requires the Quant network's QNT token to be licensed. However, governance within the consortium network is still dependent on the user group. The provided software can also include functionality for the consortium to make decisions, as Hyperledger Firefly does.

\subsection{Open Source}
Out of the nine projects examined, all but one are open source, and their code can be audited by anyone online. Overledger is the exception. The client part of the API is documented publicly, but the internal implementation of message hashing, known as TrustTag, is patented by Quant and is closed-source. While having the core part of the solution be closed-source allows Quant to gain an advantage in the current market and prevent possible exploits, it also comes with the drawback of possibly slowing down the advancement of their interoperability solution to be compatible with other blockchain technologies.

\subsection{Trust}
\emph{Trust} in a given blockchain is the ability to engage in transactions (e.g., transfer of assets) with an entity that a user has not interacted with before or does not necessarily trust.
If there is a guarantee that the entity will behave correctly, then the system is \textit{trustless}. However, if one needs to \textit{trust} the entity to behave correctly -- with no provable guarantee -- then the system is \textit{trusted}. This property is closely linked with how the verification happens on the system. In the traditional blockchain architecture, trust is established by having the nodes check each other's work and agree. Interoperability solutions that leverage a blockchain (Hub and Spoke, Decentralized Oracle) are \textit{trustless} because the blockchain maintains the system's security. However, decentralized relays need to be \textit{trusted} because they are designed to be operated by a permissioned consortium of entities.

\begin{table*}[!htbp]
\resizebox{\textwidth}{!}{%
\begin{tabular}{|c|c|c|c|c|c|c|c|c|}
\hline
 & \textbf{Summary} & \textbf{Integration Mode} & \textbf{Validation} & \textbf{Upgradability} & \textbf{Adoptability} & \textbf{Governance} & \textbf{Open Source} & \textbf{Trust} \\ \hline
\textbf{Overledger} & \begin{tabular}[c]{@{}c@{}}Message layer for \\ multi-chain applications\end{tabular} & Decentralized Relay & \cellcolor[HTML]{FCFC85}Hybrid & \cellcolor[HTML]{9FFBBE}Through plugins & \cellcolor[HTML]{9FFBBE}Allow metadata field & \cellcolor[HTML]{F79592}Off-chain & \cellcolor[HTML]{F79592}FALSE & \cellcolor[HTML]{F79592}Trusted \\ \hline
\textbf{Hyperledger Cactus} & \begin{tabular}[c]{@{}c@{}}Integration framework \\ for multi-chain transactions\end{tabular} & Decentralized Relay & \cellcolor[HTML]{F79592}Third Party & \cellcolor[HTML]{9FFBBE}Through plugins & \cellcolor[HTML]{FCFC85}Implement plugin & \cellcolor[HTML]{F79592}Off-chain & \cellcolor[HTML]{9FFBBE}TRUE & \cellcolor[HTML]{F79592}Trusted \\ \hline
\textbf{Hyperledger Firefly} & \begin{tabular}[c]{@{}c@{}}API layer for \\ multi-chain applications\end{tabular} & Decentralized Relay & \cellcolor[HTML]{F79592}Third Party & \cellcolor[HTML]{9FFBBE}Through plugins & \cellcolor[HTML]{FCFC85}Implement plugin & \cellcolor[HTML]{F79592}Off-chain & \cellcolor[HTML]{9FFBBE}TRUE & \cellcolor[HTML]{F79592}Trusted \\ \hline
\textbf{Polkadot} & \begin{tabular}[c]{@{}c@{}}Enable value and data to be \\ sent across multiple chains\end{tabular} & Hub and Spoke & \cellcolor[HTML]{9FFBBE}Native & \cellcolor[HTML]{FCFC85}No hard fork & \cellcolor[HTML]{F79592}\begin{tabular}[c]{@{}c@{}}Implement protocol/\\ build in Substrate\end{tabular} & \cellcolor[HTML]{9FFBBE}On-chain & \cellcolor[HTML]{9FFBBE}TRUE & \cellcolor[HTML]{9FFBBE}Trustless \\ \hline
\textbf{Cosmos} & \begin{tabular}[c]{@{}c@{}}Enables transfer of \\ value between chains\end{tabular} & Hub and Spoke & \cellcolor[HTML]{9FFBBE}Native & \cellcolor[HTML]{F79592}Possible hard fork & \cellcolor[HTML]{F79592}\begin{tabular}[c]{@{}c@{}}Implement protocol/\\ build in Tendermint\end{tabular} & \cellcolor[HTML]{9FFBBE}On-chain & \cellcolor[HTML]{9FFBBE}TRUE & \cellcolor[HTML]{9FFBBE}Trustless \\ \hline
\textbf{Loom} & \begin{tabular}[c]{@{}c@{}}Enable interoperability\\ for Etherum DApps\end{tabular} & Hub and Spoke & \cellcolor[HTML]{9FFBBE}Native & \cellcolor[HTML]{F79592}Possible hard fork & \cellcolor[HTML]{F79592}\begin{tabular}[c]{@{}c@{}}Loom sidechain must \\ be implemented\end{tabular} & \cellcolor[HTML]{9FFBBE}On-chain & \cellcolor[HTML]{9FFBBE}TRUE & \cellcolor[HTML]{9FFBBE}Trustless \\ \hline
\textbf{Hybrix} & \begin{tabular}[c]{@{}c@{}}Blockchain-agnostic \\ interoperability protocol\end{tabular} & Decentralized Relay & \cellcolor[HTML]{FCFC85}Hybrid & \cellcolor[HTML]{9FFBBE}Through P2P daemon & \cellcolor[HTML]{9FFBBE}Allow metadata field & \cellcolor[HTML]{F79592}Off-chain & \cellcolor[HTML]{9FFBBE}TRUE & \cellcolor[HTML]{F79592}Trusted \\ \hline
\textbf{Chainlink} & \begin{tabular}[c]{@{}c@{}}Network of Decentralized \\ Oracle Networks (DON)\end{tabular} & Oracle & \cellcolor[HTML]{9FFBBE}Native & \cellcolor[HTML]{9FFBBE}Contract can self-generate & \cellcolor[HTML]{9FFBBE}No requirement & \cellcolor[HTML]{9FFBBE}Onchain & \cellcolor[HTML]{9FFBBE}TRUE & \cellcolor[HTML]{9FFBBE}Trustless \\ \hline
\textbf{Poly Network} & \begin{tabular}[c]{@{}c@{}}Interoperability protocol for\\ heterogenous blockchains\end{tabular} & Hub and Spoke & \cellcolor[HTML]{FCFC85}Hybrid & \cellcolor[HTML]{F79592}Possible hard fork & \cellcolor[HTML]{F79592}\begin{tabular}[c]{@{}c@{}}Implement\\ Smart contract \& Relayer\end{tabular} & \cellcolor[HTML]{9FFBBE}On-chain & \cellcolor[HTML]{9FFBBE}TRUE & \cellcolor[HTML]{9FFBBE}Trustless \\ \hline
\end{tabular}%
}
\caption{Overview of Existing Blockchain Solutions.\\The cells are colored to denote the security of the properties. \colorbox[HTML]{9FFBBE}{\textbf{Green:}} Good, \colorbox[HTML]{FCFC85}{\textbf{Yellow:}} OK, \colorbox[HTML]{F79592}{\textbf{Red:}} Can be improved}
\label{tab:final_table}
\end{table*}


\section{Conclusion}

We examined nine different implementations for blockchain interoperability. These projects are compared based on properties we consider essential for such a framework. In particular, upgradability and adoptability can directly lead to the widespread adoption of an interoperability solution.
Among the discussed solutions, three broad integration modes can be seen in these projects: decentralized relays, hub and spoke, and decentralized oracles. The decentralized relay architecture is well suited for dealing with arbitrary data and can be easier to configure. However, depending on the third-party validation service, it can also be less secure. The hub and spoke architecture can provide better security and validation by using secondary blockchains to implement the system. 
However, achieving complete upgradeability can be more difficult, requiring more work for developers to join new networks to the system. The decentralized oracle approach is a tool for blockchains to use off-chain data. Although uni-directional, it can also facilitate interoperability by allowing different blockchains to read data from others. Chainlink provides a novel solution to upgradability by enabling smart contracts to generate updated copies.
While it is still too early to make any statement about which approach is better, we have highlighted some differences in these implementations to provide a bottom-up view of the current interoperability landscape in the blockchain space.


\section*{Acknowledgement}

\noindent We acknowledge the support from National Science Foundation Industry–University Cooperative Research Centers (NSF IUCRC) Center for Research toward Advancing Financial Technologies (CRAFT) research grant for this research.

\bibliographystyle{plain}
\bibliography{references}
\end{document}